\documentclass{llncs}

\usepackage{amsmath,amssymb,amscd,amsfonts}
\usepackage[pdftex]{color, graphicx}

\begin{document}

\bibliographystyle{splncs}

\title{Secure Remote Voting Using Paper Ballots}
\author{\L{}ukasz Nitschke}
\institute{
Adam Mickiewicz University, Pozna\'n, Poland
}

\maketitle

\begin{abstract}
Internet voting will probably be one of the most significant achievements of the future information society. It will have an enormous impact on the election process making it fast, reliable and inexpensive. Nonetheless, so far remote voting is considered to be very difficult, as one has to take into account susceptibility of the voter's PC to various cyber-attacks. As a result, most the research effort is put into developing protocols and machines for poll-site electronic voting. Although these solutions yield promising results, they cannot be directly adopted to Internet voting because of secure platform problem. However, the cryptographic components they utilize may be very useful. This paper presents a scheme based on combination of mixnets and homomorphic encryption borrowed from robust poll-site voting, along with techniques recommended for remote voting -- code sheets and test ballots. The protocol tries to minimize the trust put in voter's PC by making the voter responsible for manual encryption of his vote. To achieve this, the voter obtains a paper ballot that allows him to scramble the vote by performing simple operations (lookup in a table). Creation of paper ballots, as well as decryption of votes, is performed by a group of cooperating trusted servers. As a result, the scheme is characterized by strong asymmetry -- all computations are carried out on the server side. In consequence it does not require any additional hardware on the voter's side, and offers distributed trust, receipt-freeness and verifiability.
\end{abstract}

\section{Introduction}
If we take a critical look on the traditional voting methods that we have been using for years, we can observe many opportunities for fraud along with the inability of the citizens to verify the election results. This gives a strong motivation for computer scientists to design electronic mechanisms that could realize voting, and that would not only disable cheating and allow checking, but also lower the costs and increase availability. Unfortunately, such electronic solutions, contrary to traditional voting, have to face an inherent threat that any security hole may allow massive abuse (this is an exemplification of a general phenomena well described by Schneier \cite{Schneier}, and denoted as \emph{class break}). More formally electronic voting should meet the following requirements.
\begin{itemize}
  \item Anonymity, privacy -- voters' choice should remain their secret.
  \item Receipt-freeness -- voter should be unable to convince a third party of the vote decision and, as a consequence, should be unable to sell his or her vote. This property is also achieved if the voter has effective measures of deceiving a potential buyer.
  \item Verifiability -- voter should be able to check correctness of every stage of the protocol. He or she should be empowered to verify the tallying process (global verifiability) and check if his or her vote was included (individual verifiability). Usually individual verifiability requires a lookup in a public catalog, whereas, global verifiability demands performing some computations. This implies the fact that an average voter delegates global verifiability to experts or watchdog organizations.
\end{itemize}

Three approaches to the problem of electronic voting have been proposed so far \cite{CIVTF00} \cite{Oppliger02}.
\begin{itemize}
  \item \emph{Poll-site voting} also denoted as DRE (Direct Recording by Electronics) -- special voting machines with dedicated software are installed in voting booths at polling stations. Voters can cast votes by interacting with such a machine, and in some cases he or she can receive a receipt for verification. The terminal and the environment can be controlled. Moreover, some steps of the protocol may by performed by an election official, for instance the voter can be personally authorized.
  \item \emph{Kiosk voting} --  voting takes place through publicly available terminals (e.g. sophisticated ATMs or dedicated state-owned machines). In this scenario only the terminal can be controlled.
  \item \emph{Voting via Internet} performed by a client-server application, run by voter's PC/mobile phone/PDA/smart card, and on the server side, by trusted authority or authorities. Neither the terminal not the environment can be controlled.
\end{itemize}

\paragraph{Poll-site voting.} Recently, the effort of researchers is mostly put into developing protocols and machines for poll-site voting. Among the listed approaches, this is the least demanding from the security perspective, as we can assume control of the voter, and on the environment (the officials are present at the poll-site). Several well designed solutions have been proposed so far, including: \cite{Chaum04} \cite{Neff01} \cite{Adida06} \cite{RS06}. The machines used at polling stations produce encrypted ballots. The verification of the encryption process takes place through printing two ballots and letting the voter choose which to check. The chosen ballot is compromised by providing (usually printing) additional data which allows verification. The process requires a computer program and equipment (e.g. scanner). In practice, the verification is meant to be carried out by watchdog  organizations that collect the ballots at polling stations or somewhere else. As a consequence, the remaining ballot is believed to be properly constructed and is used to cast a vote. The ballot without additional data cannot serve as an evidence for vote selling purposes (receipt-freeness). Encrypted votes are published and processed by election authorities to obtain the final result. Every voter can check if his or her encrypted vote participated. Correctness of the processing can be universally verified. Interactive testing of machine encryption plus verification of the processing give strong assurance that cheating is impossible. Although verifiability and receipt-freeness were successfully adopted to poll-site voting, they cannot be easily applied to Internet voting by employing the same techniques (interactive testing). This is the consequence of the following observations. 
\begin{itemize} 
  \item We cannot replace dedicated machines with voters' personal computers. This is caused by the fact that a PC is not tamper-proof, and may leak secret information to the voter, enabling vote selling. For instance, interactive checking of ballots relies on the assumption that the voter does not learn verification data for the proper ballot.
  \item It is difficult and expensive to reduce the size of voting machines used in poll-site voting to make them portable (integrated printer and other implementation aspects).
  \item Verification of ballots requires either a trusted testing device or a third party organization. 
\end{itemize}

\paragraph{Remote voting.} The remote voting approach is the most convenient and cost-effective. It also reflects the needs of the modern society. Nonetheless, it is considerably  more challenging, as we have to take into account various cyber-attacks (possibly launched from a hostile country), and less control of the voter \cite{Rubin01} \cite{Oppliger02}. Vulnerabilities of voter's PC may open the door to many serious abuses, e.g. automated vote selling or malicious changing of votes. Several countermeasures have been proposed to minimize the trust put in voter PC. Apart from the expensive ones (trusted hardware) and the idealistic ones (clean operating system) code sheets, and test ballots seem to be promising \cite{Oppliger02}. Code sheets impose a complete asymmetry in the computational sense -- no computations are done on the voter's side. This is achieved by providing voters with ballots that contain unique codes representing candidates (different set of codes for each ballot). Each candidate code has a verification code assigned to it. The PC is used to pass the entered code on to the election authority, which returns the relevant verification code. The response is displayed by the PC, integrity of remotely cast vote. Honest election authority may prevent cyber-attacks, but a dishonest one can try to influence elections results or breach voter's privacy. An exemplification of code sheets scheme is SureVote system \cite{Chaum01}. Test ballots is an approach that suggests introducing special ballots in the voting stage. The ballots should be unrecognizable for the tallying authorities, so that they are unable to predict for which ballots they will have to reveal processing after publication of the result.
In addition to the SureVote system, there are attempts to solve the untrusted platform problem by utilizing trusted hardware (\cite{LK02}), but these solutions have to face a serious threat of malicious producers and kleptography \cite{YY96}. 

\section{Sketch of the new protocol}
\subsection{Design goals}
Design of the presented protocol was motivated by four main aims.
\begin{itemize}
  \item Providing verification and receipt-freeness at the same time.
  \item Easy and inexpensive integration with traditional elections. It means that no additional hardware on the client side is required and that remote voters may be recognized before traditional elections start to prevent double voting. 
  \item Reduction of trust put in voter's PC and software it runs, by assuring that voter's computer is unable to record and change choices made by its user (neither randomly nor intentionally).
  \item Distribution of trust put in authorities. 
\end{itemize}

The main idea of the protocol is that the voter encrypts his vote manually by performing simple operations on his paper ballot. This assumption creates a need for a scrambling method that is feasible to perform on a piece of paper. Permutations, or cyclic rotations in particular, lend themselves naturally to this goal. Operations performed by the voter can be inverted by a group of cooperating servers that perform distributed computations. This is how the trust is  distributed, and contrary to code sheets, the process of decrypting votes can be publicly controlled. This model also includes distributed creation of paper ballots. The ballot can be interactively tested using similar techniques as in poll-site machines.

\subsection{Actors}
Apart from the voters, the protocol employs the following trusted authorities. 
\begin{itemize}
  \item $EC_1$ (\emph{Election Committee 1}) -– provides an on-line voting service, whose users are authorized using known protection mechanisms. Communication with $EC_1$ can be established through an authenticated (in both directions), private channel. 
  \item $EC_2$ (\emph{Election Committee 2}) -– prints paper ballots used to encrypt user choice.
  \item $A_1 , A_3 ,..., A_{\lambda}$ -- authorities that participate in the process of creating paper ballots, and in the process of decrypting votes, they assure also distribution of trust and audit.
  \item $BB$ -- bulletin board, provides an authenticated public channel -- allows to publish and then access signed messages.
\end{itemize}

\subsection{The protocol from the voter's perspective}
\begin{enumerate}
  \item\emph{(Registration)} Registration procedure should be similar to applying for an electronic bank account. A citizen fills in an application form and submits it personally to the local administration office where he is personally authorized (based on his signature and ID card). After a reasonable period of time the voter is able to receive his elections kit from $EC_1$. The kit contains credentials which enable remote authentication of the user. The methods used here may be similar to the ones used in e-banking, e.g. PIN, password, one-time passwords, token. A token integrated with electronic ID card or signatures-enabled ID is considered to be an optimal solution.
  \item\emph{(Obtaining of paper ballot)} A few months before elections a citizen, who is willing to cast his vote via Internet is obliged to visit the local administration office in order to obtain a paper ballot, and to be personally authorized. It is assumed that the voter has already gained access to the election service (see: registration). The voter chooses two ballots, then decides which one should be verified. The selected ballot is recoded by an election official, and can be verified by the voter or by a civic committee. The other ballot is separated from the part containing validation data, and serves as proper means of casting a vote. The validation data is destroyed in the presence of the voter.

\item\emph{(Manual encryption)} Every ballot has a simple operation assigned to it. The transformation is represented by a table so that it is easy for the voter to encode his or her candidate number $v$ by the value $v + sh \mod c$. The voter marks the candidate's number and reads the underlying encryption of the vote.  
  \item\emph{(Vote submission)} The voter log into the voting system, enters $id$ and the encrypted vote.
  \item\emph{(Publication of encrypted votes)} $EC_1$ publishes voters' names along with encrypted votes on the $BB$.
  \item\emph{(Verification)} Voter checks if his vote reached bulletin board in an unchanged form. 
  \item\emph{(Vote results publication)} The votes are decrypted by trusted authorities $A_1, A_3,\ldots,A_{\lambda}$ and published.
\end{enumerate}

\section{Building blocks}
Most of the building blocks we employ are based on the ElGamal public key cryptosystem. Let $p$ be a large prime, $g$ a generator in $\mathbb{Z}_p$ and $x$ a random element of $\mathbb{Z}_p$. We define: ElGamal private-public key pair as $(x; p, g, y(= g^{x} \mod p))$, ElGamal encryption function $e_y(m) = (m y^{k}, g^{k})$ and decryption function $d_x(a,b) = a / b^x$
Owner $A$ of asymmetric keys $(x; p,g,y)$ can prove non-interactively that a given cipertext $(a, b)$ is an encryptption of message $m$ using a zero-knolege proof of equality of discete logs ($log_{b}(a/m) = log_{g}(y)$) \cite{CP93}. We will denote the proof as $nizk(m,(a,b))$.

\subsection{Mix Networks}
One of the most important branches in research of electronic voting are protocols based on mix networks. Mix network protocols allow to shuffle a list of encrypted messages in a distributed way by $\lambda$ trusted parties (mix servers). Each party $A_i$ sequentially permutes and transforms elements on the list. The resulting list is passed on to the next mix server via an authenticated public channel ($BB$). Transformations carried out by a single server obfuscate relations between input and output elements. Therefore, it is hard to determine the secret permutation of a single sever, and in consequence the global permutation of the whole mixnet.  
We need two functions to perform distributed shuffling:
\begin{itemize}
     \item $o(m)$ -- creates an initial encrypted form (called onion) of $m$ that passes through the mix servers
    \item $t_{k}(c)$ -- transforms cipher text $c$ into $c'$ so that: $c'$ encrypts the   same message as $c$, and it is difficult to prove this fact without the knowledge of the randomizing value $k$.
\end{itemize}

There are different types of mixnets, depending on the transformation function. We will employ partially decrypting mixnets . This type of mixnet is characterized by the fact that each server partially decrypts elements on its input list and the last server yields messages. Such a mixnet based on ElGamal can be build by defining the $o$ and $t$ functions as follows \cite{PIK93}: $o_{k}(m) = e_{y_{1} y_{2} ... y_{\lambda}}(m)$, $t_{k}((a,b)) = (a (y_{i+1} y_{i+2} ... y_{\lambda})^{k} / b^{x_i}, b \cdot g^{k})$. Where $(x_{i} ,y_{i})$ are $A_i$ asymmetric keys.

\subsubsection{Randomized Partial Checking}
A mix network protocol can be employed as a component of electronic voting if it can be guaranteed that none of the list elements was replaced or maliciously altered. This property called \emph{robustness} is provided by additional checking. Randomized Partial Checking is a fairly simple and effective verifying technique which was introduced in \cite{JJR02}. The mix servers are obliged to reveal a random half of their input-output relations, with the assurance that no path of length greater than 2 can be uncovered. To achieve this property the servers are paired, and forced to uncover complementary halves of their transformations. 

\noindent In more detail, RPC consists of the following steps:
\begin{enumerate}
  \item \emph{(Before shuffling)} The mix servers publish commitments to their permutations ($pcommit(\pi) = (commit(\pi(1)),...,commit(\pi(n)))$).
  \item \emph{(After shuffling)} The servers establish a fairly chosen value $r = r_{1}\oplus r_{2}\oplus ...\oplus r_{\lambda }$ -- each server contributes its $r_i$ using commitments, so that no party is able to determine $r$. Then a value $q = hash(r, content(BB))$ is computed, and $q_i = hash(q, i)$ are derived. Values $q_i$ determine transitions to be revealed in pair $i$. To prove validity of a selected transition of $j$-th input server $A_i$ publishes a value $validator(i,j)$ that may consist of $decommit(\Pi_i(j)), k_{ij}$, where $k_{ij}$ is the randomization value used in the $j$-th transformation.
\end{enumerate}

\subsection{Homomorphic Encryption and Re-encryption}
The ElGamal scheme has the property that having two encrypted messages, one can calculate the ciphertext of  multiplication of the two messages. This can be achieved by simply multiplying the two ciphertexts. This property is known as $(\cdot{},\cdot)${}-homomorphism. For the sake of this paper $(\cdot{},+)${}-homomorphism is more useful. However, this requires a small modification of the original ElGamal.
\begin{itemize}
  \item  $he_{y}(m_{1})=(h^{m_{1}}\cdot y^{k_{1}}, g^{k_{1}})$, $he(m_{2})=(h^{m_{2}}\cdot y^{k_{2}}, g^{k_{2}})$ 
  \item  $he_{y}(m_{1})\cdot he_{y}(m_{2})=(h^{m_{1}}\cdot y^{k_{1}}\cdot h^{m_{2}}\cdot y^{k_{2}}, g^{k_{1}}\cdot g^{k_{2}})=(h^{m_{1}+m_{2}}\cdot y^{k}, g^{k})=he_{y}(m_{1}+m_{2})$ 
\end{itemize}
$h^m$ is obtained from $he_{y}(m)$ by performing regular ElGamal decryption, then $m$ is found through exhaustive search or lookup in a precomputed table. Note, that if we take $p$ such that a large prime $q | p-1$ and a small $r | p-1$ then assuming that $ord(g) = q$ and $ord(h) = r$ we can perform encrypted additions in $\mathbb{Z}_r$.

\subsection{Computing Mixnet}
If we combine mixnets with the idea of homomorphic encryption we obtain a protocol for distributed computation that has the property that it obfuscates the relations between input and output values. Computations performed by such a network can be used to anonymously invert the adding operation (cyclic rotation) performed by the voter. We now obtain two new $o(\cdot)$, $t(\cdot)$ functions:
\begin{itemize}
    \item $ho_{k}(m) = he_{y_{1} y_{2} ... y_{\lambda}}(m)$,
    \item $ht_{k,l}((a,b)) = ( a \cdot h^{l} \cdot (y_{i+1} y_{i+2} ... y_{\lambda})^{k} / b^{x_i}, b \cdot g^{k})$, $l$ is a value added in a given transformation. 
\end{itemize}

\section{The Protocol}
\subsection{Setup}
\noindent Notation: $n$ -- number of paper ballots, $c$ -- number of candidates; $p$ -- secure, public prime, such that a large prime $q | p-1$ and $c | p-1$, $g,h$ -- generators in $\mathbb{Z}_p$ of order $q$, $c$ respectively; $(x_i; p, g, y_i)$ -- $A_i$ asymmetric keys; $(x_{EC_2}; p, g, y_{EC_2})$ -- $EC_2$ asymmetric key. Before elections start the following steps need to be fulfilled.
\begin{enumerate}
  \item $EC_1$ chooses a permutation $\pi_0 : \mathbb{Z}_n \rightarrow \mathbb{Z}_n$, and publishes the commitment. $$EC_1 \longrightarrow BB : pcommit(\pi_0)$$ 
  \item Each $A_i$:
  \begin{enumerate}
    \item chooses a permutation $\pi_i : \mathbb{Z}_n \rightarrow \mathbb{Z}_n$, and a vector of small integers ($l_{ij} < c$): $l_i = (l_{i,1}, l_{i,1},..., l_{i,n})$;
    \item publishes the commitment to its permutation and to the values $l_{ij}$ $$A_i \longrightarrow BB : pcommit(\pi_i), commit(l_i).$$
  \end{enumerate}
\end{enumerate}
  
\subsection{Actions of the protocol}
\paragraph{Creation of ballots.} Creation of paper ballots involves sending $n$ pairs of partially decrypting onions through the mixnet. The first onion in a pair carries an identifier of the input position, while the second one uses homomorphic encryption to accumulate the sum $sh$ that forms the cyclic rotation printed on the paper ballot. 
  $$c_{0, j} = e_{y_{1}y_{2}...y_{\lambda}y_{EC_2}}(\pi_0(j)), hc_{0, j} = (1, 1), j = 1,..,n$$
  $$EC_1 \longrightarrow BB : (c_{0,j}), (hc_{0,j})$$
The pairs of onions are then being processed by the authorities $A_i$ ($c_{i, j} = t_{k_{i,j}, l_{i,j}}(hc_{i-1, j})$, $hc_{i, j} = ht_{k_{i,j}, l_{i,j}}(hc_{i-1, j})$), $i = 1,2,...,\lambda-1$ and passed on to the next authority through the bulletin board. Note that the resulting identifiers and values $sh$ remain secret to the public audience, as they are encrypted with $EC_2$ public key.

\paragraph{Distribution and checking of ballots.} Each voter $V$ personally obtains two paper ballots. He or she chooses one for verification and learns its validation values. The ballot identifier is scanned by an election official and marked as invalid by $EC_2$ on the $BB$. The values it contains can be verified: $\hat{id}$, $\hat{sh}$, $\hat{c}_{\lambda, k}, \hat{hc}_{\lambda, k}$ (relevant output onions), $nizk(\hat{id}, \hat{c}_{\lambda, k}), nizk(h^{\hat{v}}, \hat{hc}_{\lambda, k})$ (proofs). The other ballot provides the voter with $id, sh$, which are used for voting. 

\paragraph{Casting votes.} Each voter $V$ encrypts his vote $v$ obtaining $ev = v + sh \mod c$. Encrypted vote is sent along with $id$ to $EC_1$ through an authenticated channel.
       $$V \stackrel{auth}\longrightarrow EC_1 : id, ev$$
The election authority publishes position on the input list $p = \pi_{0}^{-1}(id)$, voter's identifier, his encrypted vote, the onion $hc'_{0, p} = ho_{k'_{0,j}}(ev)$, and $k'_{0,j}$ as proof of the correctness of the onion.
       $$EC_1 \longrightarrow BB : p, V, ev, hc'_{0, p}, k'_{0,j}$$    

\paragraph{Recovering and counting votes.}
The encrypted votes that have been available on the bulletin board enter the same mixnet, that uses the same $l_{ij}$ values, but instead of adding they are now subtracted by $A_i$ ($hc'_{\pi_i(j)} = ht_{k'_{i,j}, -l_{i,j}}(hc'_{i-1, j})$). The votes are made available on the $BB$. The onions that reach positions on the output list marked as invalid are traced back.
  
\subsection{Mix-and-compute verification}
For verification of the two stage process of creating ballots and recovering votes we need also a two stage validation technique. Splitting RPC directly into two phases -- revealing $1/4$ of transitions twice -- does not work. This is because before the second stage of testing a malicious mix server would be able to pinpoint transitions that cannot be tested according to the rule that in a pair of servers no path of length two can be uncovered. Therefore, we propose a different two stage version of RPC, in which the servers are grouped in 4-tuples consisting of two pairs.
\begin{enumerate}
  \item \emph{(After creation of ballots)} $1/4$ of transitions of each server is revealed preserving paths in uncovered pairs.

  \item \emph{(After recovery of votes)} Within each 4-tuple one pair is selected to reveal remaining the $1/4$ of mappings in the RPC fashion. The servers in the other pair reveal transformations independently.  

\end{enumerate}
Now the probability that replacement of $n$ onions will remain unnoticed is $(1/6)^n$. The transitions selected to be verified are determined in a way similar to regular RPC. However, the validating values also include elements of sums -- $l_{ij}$. Transitions chosen to be revealed are uncovered in both stages.

\section{Further Enhancements and Remarks}
\paragraph{Receipt-freeness and verifiability.} Providing verifiability and receipt-freeness (inability to sell votes) is the biggest challenge in design of voting protocols. In our protocol the voter is given two paper ballots. He or she chooses one of them to reveal its validation values. The ballot is marked as compromised and can be verified by a watchdog organization. As a result the voter believes that the other ballot, whose validation values were destroyed, is also valid. But he or she is unable to prove it to anybody else (who was not present during interactive testing), and sell the vote. Since the process of ballot creation and recovery of votes is controlled, the voter is assured that his vote was properly counted.  

\paragraph{Test ballots and interactive testing.} 
The basic protocol presented above shows verification of mixing operations. Nonetheless, there are two steps of the protocol that deserve extraordinary suspiciousness -- printing of ballots, and putting encrypted votes on the input list before decryption. Printing is verified be interactive testing -- choosing one ballot out of two for thorough investigation. The output position (from the ballot creating mixing) is then marked as compromised. An interesting idea is to utilize the compromised ballots as test ballots. They might be introduced into the mixnet by verification organizations or voters themselves to strengthen verification of the decryption process (and the input list). The decrypted votes that hit compromised positions can be traced back, and voters who decided to check could verify their test ballot. In this sense test ballots are a real trust increasing factor.

\paragraph{Code sheets.} The scheme would certainly benefit form an immediate assurance that votes cast by the voters reached the authority unchanged. To achieve this goal we can employ verification codes inserted in the identifier onion by $EC_1$ during creation of ballots. Natural candidates for the codes are truncated digital signatures of $EC_1$. 

\paragraph{Colluding authorities. } The presented protocol assumes that the authorities $EC_1$ and $EC_2$ are in a conflict of interest -- for instance they are controlled by the ruling and opposition party. Otherwise, they would be able to violate users privacy and try to introduce fake votes. This is caused by the fact that one authority ($EC_1$) controls the input to the computing mixnet while the other party ($EC_2$) controls the output.

\paragraph{K-out-of-L voting.} In the basic setting the proposed scheme offers 1-out-of-L voting, which means that we can choose only one candidate. However, we can easily extend it by adding multiple cyclic rotations to K-out-of-L or K-out-of-L-ordered voting. This requires increased number of homomorphic onions processed by the authorities. 

\section{Conclusions}
So far, no viable solution to the problem of remote electronic voting has been proposed. The computational model presented in this paper is a step towards overcoming vulnerabilities of operating systems, personal computers and the Internet. The solution also offers receipt-freeness, and full verification of every step. Crucial parts of the verification can be carried out by an average voter while the more complicated procedures may be delegated to experts or independent organizations. We also showed that autonomously computing mix servers may be a useful component of cryptographic protocols.

\end{document}